# Three Comments on "A Simple Incremental Modelling of Granular-Media Mechanics"

## P. Evesque

Lab MSSMat, UMR 8579 CNRS, Ecole Centrale Paris
92295 CHATENAY-MALABRY, France, e-mail evesque@mssmat.ecp.fr

**Abstract:**
*Using the recent incremental modelling, it is shown that the trajectory of a sample in the phase space of soil mechanics in the vicinity of the critical state is not governed by the rigidity matrix, but by its variations. The characteristics are used to predict that Young modulus tends to 0 as $\sigma_1/\sigma_2$-M-1 near the critical point. An attempt to understand and predict unstable behaviours is done using the same modelling. Compatibility of this modelling with results on soil liquefaction is emphasised.*
_______________________________________________________________________

**Recall:** In a recent paper [1-3], the author proposed a simple non-linear incremental modelling to describe the mechanics of granular materials. At a given working point, characterised by the stress field $\sigma$, the mean density and the anisotropy, this modelling assumes a linear relationship (Eq. 1) between the stress increments $\delta\varepsilon$ and the strain increments $\delta\sigma$ in a zone which is large enough to contain all axial compression tests. It has been argued that the hypothesis of an isotropic response ($\alpha=1, \nu=\nu'=\nu''$) is sufficient in many cases, as far as one assumes that $\nu$ depends on $\sigma_1/\sigma_3$ only (i.e. $\nu = 1 - \sigma_1/\{2(1+M)\sigma_2\}$, where $M=2\sin\varphi/(1-\sin\varphi)$, with $\varphi$ being the friction angle) and that $C_o$ varies with deformation. It has been argued also that this isotropic assumption is no more valid in the vicinity of the critical state. Anyhow, this modelling is non linear since the pseudo-Poisson coefficient $\nu$ and the pseudo-Young modulus $1/C_o$ both evolve.

$$\begin{pmatrix} d\varepsilon_1 \\ d\varepsilon_2 \\ d\varepsilon_3 \end{pmatrix} = -C_o \begin{pmatrix} 1 & -\nu' & -\nu' \\ -\nu & a & -\nu'' \\ -\nu & -\nu'' & a \end{pmatrix} \begin{pmatrix} d\sigma_1 \\ d\sigma_2 \\ d\sigma_3 \end{pmatrix} \qquad (1)$$

In this approach, the characteristic states are defined as the states for which no volume change is obtained after an increment of stress $d\sigma_1$, and the critical states are those characteristic states which are obtained after very large deformation when no volume change is observed any more. These states are observed experimentally when $\sigma_1/\sigma_2=1+M$. This imposes that $\nu=1/2$ and $\alpha-\nu'-\nu''=0$ when $\sigma_1/\sigma_2=1+M$.

**Remark 1: In the vicinity of the critical point**, $\nu$ tends to 1/2 but the pseudo-Young modulus $1/C_o$ tends to 0. This is not the case for other characteristic states, which exhibit $C_o \neq 0$. So, near the critical state, the volume variation can be written as:

$$d\varepsilon_v = -C_o\{(1-2\nu)d\sigma_1 + 2(\alpha-\nu'-\nu'')d\sigma_2\} \qquad (2)$$

Now, experimental trajectories in the ($q=\sigma_1-\sigma_2$, $p=(\sigma_1+\sigma_2+\sigma_3)/3$, $v$) space which arrive at the critical points and are continuous from both left hand side and right hand side; but $C_o$ tends to infinity at this point; so, Eq. (2) imposes that $1-2\nu$ and $(\alpha-\nu-\nu'')$ tend also to 0 when the system tends to the critical point. This imposes 2 conditions:

$$C_o(1-2\nu)=A \qquad (3a)$$

$$C_o(\alpha-\nu'-\nu'')=B \qquad (3b)$$

with A and B being finite; so, combining Eq. (3a) with the Rowe's relation $(1-2\nu)=[\sigma_2(1+M)-\sigma_1]/\sigma_2$, which controls the variation of $\nu$, leads to the pseudo-Young modulus $1/C_o$ to be:

$$1/C_o = E_{\text{vicinity of critical state}} = [\sigma_2(1+M)-\sigma_1]/(A\sigma_2) \qquad (4a)$$

and Eq. (3b) imposes in turn that:

$$(\alpha-\nu'-\nu'')=B(1-2\nu)/A=B[\sigma_2(1+M)-\sigma_1]/(A\sigma_2) \qquad (4b)$$

This fixes the condition of the divergence of $1/C_o$ near the critical state.

So, the direction of the trajectories near the critical state is not controlled by the rigidity matrix as it was said in the Hvorslev-Roscoe section of [1] (p.8), but by its variations. However, this seems to have little consequence and does not seem to invalidate the results of [4].

**Remark 2: unstable behaviour of materials submitted to $\sigma_1/\sigma_2>1+M$**: It is known from elasticity [5] that materials with Poisson coefficient $\nu$ larger than 1/2 are not physically stable since the shape of their free energy surface would exhibit a saddle point instead of a minimum, according to:

$$F = F_o + \varepsilon_{ii}^2\, E\nu/\{2(1-2\nu)(1+\nu)\} + \varepsilon_{ik}^2\, E/\{2(1+\nu)\} \qquad (5)$$

with $\nu>1/2$ and where F is the free energy and E the Young modulus. Is it the same for a granular material?

Indeed not: It is true that a granular material submitted to a stress field ratio larger than the one corresponding to the characteristic state ($\sigma_1/\sigma_2=M+1$) exhibits a pseudo-Poisson coefficient larger than 1/2. However, according to Eq.(5), the unstable process which is involved in the instability of an elastic material with $\nu>1/2$ corresponds to a volume decrease as soon as the mean stress decreases; so, consider such a material submitted to a given stress $\sigma$, and consider some time fluctuation $\delta\sigma<0$, this $\delta\sigma$ will generate a $\delta v<0$ which will release stress $\delta\sigma'$ so that $\delta\sigma$ will be amplified leading to the instability of the configuration.

In the case of a granular material on the contrary, linearity is valid only in a zone, so that $\nu>1/2$ when the stress ratio $\sigma_1/\sigma_2$ increases; so it is valid in the case of a compression only; so, the mechanism of instability cannot apply to an expansion ($\delta\sigma<0$), and instability cannot develop there.

Anyhow, it is well known that instability occurs in granular materials; for instance, it is obvious that unstable behaviour can be found with $C_o$<0 during a compression test. But it has been observed from experiment that localisation of deformation can occur before $C_o$ becomes negative. This can be explained by introducing anisotropy and off-diagonal deformation [6]. Furthermore, one can see from Eq. (5) that non diagonal deformation are important too. Our simple modelling does not take into account such processes, which should be introduced for a complete discussion. It seems for instance that processes of strain rotation play an important part during the formation of strain localisation and that they control mainly the direction of localisation bands [7]. The reader can get more details in [6,7].

**Remark 3: Compatibility of the modelling with soil liquefaction:** The classical explanation for soil liquefaction has been simplified by Luong [8] by introducing the concept of characteristic line q=M'p (see Fig. 1), with q=$\sigma_1$-$\sigma_2$ , p=($\sigma_1$+$\sigma_2$+$\sigma_3$)/3 . In the present modelling, the equation of the characteristic line becomes q=M$\sigma_2$=M$\sigma_3$. This line (q=M$\sigma_2$) separates the stress domain into a contractant domain and a dilatant domain when performing a compression, as it is sketched on Figure 1; these domains are called subcharacteristic and supercharacteristic respectively.

Figure 1 caption explains how the modelling proposed in [1-3] is completely compatible with the classical approach of soil mechanics [9]. It explains also how to extend the incremental modelling of [1-3] to cyclic loading, at least qualitatively. It demonstrates that one shall distinguish between two simple experimental cases labelled case (i) and case (ii), which correspond to cycles running at $\sigma_2$=c$^{ste}$ (case i), and at constant volume (case ii). This caption describes the behaviour of materials predicted by the modelling of [1-3] in both cases (i) and (ii) as a function of the stress ratio $\sigma_1$/$\sigma_2$, i.e. in both the sub- and the super- characteristic regions; they agree qualitatively with experimental observation and classical interpretation [9].

**Miscellanous 1: undrained tests:** During an undrained test (v=c$^{ste}$), the modelling propsed here [1] predicts that the system evolves till it reaches the critical state, where its evolution stops. This is not found using classical cam-clay, but observed experimentally.

**Miscellanous 2:** this modelling, in its isotropic form, requires the knowledge of the friction angle and of the evolution of $C_o$ only; both quantities can be determined experimentally from uniaxial compression at $\sigma_2$=$\sigma_3$=constant.

On the contrary, even if we omit the models based on neural network, which are parameters consuming, most of the other models which are able to describe the same experimental results use a much larger number of parameters varying between (4 to 36), whose evolutions have to be determined. The determination of such a large number of parameters is difficult from experiments only; and it is perhaps meaningless since the unicity of the set is probably not warranted. Furthermore, these models are rather intricate so that the prediction of the behaviours becomes impossible without the help of a computer.

**Miscellanous 3:** Eq. (7) of [1] shall be read:

$$-3d\varepsilon_v/C_o = 3(1+2\alpha-2\nu-2\nu'-2\nu'')dp + 2(1-\alpha-2\nu+\nu'+\nu'')dq = 0 \qquad (7)$$

This equation fixes the variation of the tangent $dq/dp$ of the trajectory of an undrained test.

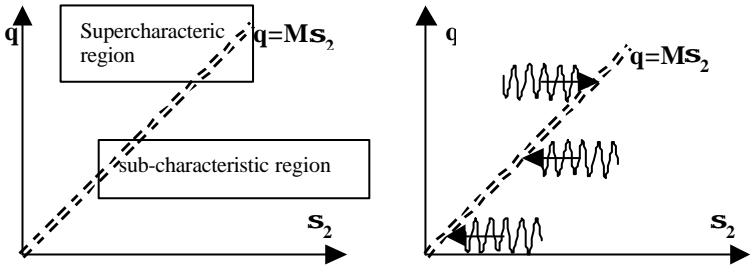

**Figure 1:** In the proposed modelling the $q=M\sigma_2$ characteristic line separate the domain where the pseudo-Poison coefficient $\nu$ during a compression is larger than 1/2 and where it is smaller than 1/2. These two domains are called the super- and sub- characteristic domains respectively and the material is there either dilatant or contractant respectively under compression. During cyclic loading the material exhibits hysteresis since loading generates irreversible deformations much larger than the one obtained during the next unloading. The hysteretic behaviour after a stress cycle shall be approximately that one obtained during the compression. Now two cases are to be considered: (i) the case when material can change of volume, and (ii) the case when it cannot change of volume. The last case (ii) corresponds to natural cases when the material is saturated with water and when its permeability is quite small so that water cannot goes out from it. This case may lead to soil liquefaction in some cases as we will see.

Let us consider case (i) first (drained materials). Applying our modelling predicts that cyclic loading shall lead to irreversible contraction and densification when stress is in the subcharacteristic domain (case i-a) ; on the contrary, it shall lead to and irreversible dilation and loosening of the material in the supercharacteristic domain (i-b). As the supercharacteristic domain can only be reached for dense enough material and as the material loosen under cycles, one shall expect that the material becomes unstable after enough cycles performed in the supercharacteristic domain (i-b). This is observed indeed experimentally. On the other hand, cyclic loading in the subcharacteristic domain lead to compaction, and the material becomes more stable (i-a); these are also two well recognised experimental facts.

In case (ii) the volume is imposed constant. So, during a stress cycle performed in the subcharacteristic (supercharacteristic) domain, the grain skeleton tends to shrink (dilate); but this becomes forbidden due to boundary condition i.e. by the presence of water; so, water tries to go in (out), which imposes in turn that local water pressure inside the pores tends to decrease (increase) reducing (increasing) the stress $\sigma_2$, if only total stresses are imposed. So cycles imposed at constant volumes impose the mean stress $\sigma_2$ to decrease (increase) in the subcharacteristic (supercharacteristic) domain as it is indicated in Fig. (1b). This process stops when the mean stress field arrives at the characteristic line, which is an attractor of this dynamics. It is worth noting that when q remains small or becomes alternatively positive and negative ($q < \delta q$, alternate cyclic path) the stress $\sigma_2$ tends to 0 after enough cycles so that the material cannot support important shear stress anymore; in this case, it deforms importantly during shearing and it becomes fluid-like. This is the well-known liquefaction process. Such liquefaction are commonly obtained during earthquakes; earthquake generates also compaction of granular material which leads to a massive water ejection generating craters.

*Acknowledgement*: Discussion with Prof. M.A. Koenders has been appreciated.

The electronic arXiv.org version of this paper has been settled during a stay at the Kavli Institute of Theoretical Physics of the University of California at Santa Barbara (KITP-UCSB), in june 2005, supported in part by the National Science Fundation under Grant n° PHY99-07949.

*Poudres & Grains* can be found at :
http://www.mssmat.ecp.fr/rubrique.php3?id_rubrique=402